
\input phyzzx
\input tables
\referenceminspace=10pc

\def\to{\rightarrow}

\def\rchi{\raise2pt\hbox{$\chi$}}

\def\psl{p\hskip-6pt/}

\def\Re{{\cal R \mskip-4mu \lower.1ex \hbox{\it e}}}
\def\Im{{\cal I \mskip-5mu \lower.1ex \hbox{\it m}}}

\def\etal{{\it et~al}.}
\def\eg{{\it e.g.}}

\def\and{\! + \!}
\def\rlh{\scriptstyle{\rightharpoonup\hskip-8pt{\leftharpoondown}}}

\def\Im{{\cal I}\mskip-5mu\lower.1ex\hbox{\it m}}
\def\harpar{\partial\hskip-8pt\raise9pt\hbox{$\rlh$}}
\def\crc{\crcr\noalign{\vskip -6pt}}
\newtoks\Pubnumtwo
\newtoks\Pubnumthree
\catcode`@=11
\def\p@bblock{\begingroup\tabskip=\hsize minus\hsize
   \baselineskip=1.5\ht\strutbox\topspace-2\baselineskip
   \halign to \hsize{\strut ##\hfil\tabskip=0pt\crcr
   \the\Pubnum\cr  \the\Pubnumtwo\cr \the \Pubnumthree\cr
   \the\date\cr \the\pubtype\cr}\endgroup}
\catcode`@=12
\Pubnum{}
\Pubnumtwo{\bf FSU-HEP-920225}
\Pubnumthree{\bf MAD/PH/692}
\date={February 1992}
\pubtype={}
\titlepage
\doublespace
\title{\fourteenrm Probing the $WW\gamma$ Vertex in $e^\pm p\to\nu\gamma
X$}
\author{\fourteenrm U.~Baur}
\vskip 2.mm
\centerline{\it Physics Department, Florida State University}
\centerline{\it Tallahassee, FL 32306, USA}
\vskip .5cm
\centerline{and}
\vskip .5cm
\author{\fourteenrm M. A. Doncheski}
\vskip 2.mm
\centerline{\it Physics Department, University of Wisconsin}
\centerline{\it Madison, WI 53706, USA}
\normalspace
\abstract
We study the prospects of testing the $WW\gamma$ vertex in
$e^- p\to\nu\gamma X$ and $e^+ p\to\nu\gamma X$ at HERA and LEP/LHC.
Destructive interference effects between the Standard Model and the
anomalous contributions to the amplitude severely limit the sensitivity
of both processes to non-standard $WW\gamma$ couplings. Sensitivity
limits for the anomalous $WW\gamma$ couplings $\kappa$ and $\lambda$ at
HERA and LEP/LHC are derived, taking into account experimental cuts
and uncertainties, and the form factor behaviour of nonstandard couplings.
These limits are found to be significantly weaker than
those which can be expected from other collider processes within the
next few years. At HERA, they are comparable to bounds obtained from
$S$-matrix unitarity.
\endpage

\def\PL #1 #2 #3 {Phys. Lett.~{\bf#1}, #2 (#3)}
\def\NP #1 #2 #3 {Nucl. Phys.~{\bf#1}, #2 (#3)}
\def\ZP #1 #2 #3 {Z.~Phys.~{\bf#1}, #2 (#3)}
\def\PR #1 #2 #3 {Phys. Rev.~{\bf#1}, #2 (#3)}
\def\PRD #1 #2 #3 {Phys. Rev.~D {\bf#1}, #2 (#3)}
\def\PP #1 #2 #3 {Phys. Rep.~{\bf#1}, #2 (#3)}
\def\PRL #1 #2 #3 {Phys. Rev.~Lett.~{\bf#1}, #2 (#3)}
%
\REF\Sam{K.~O.~Mikaelian, \PR D17 750 1978 ;\nextline
K.~O.~Mikaelian, M.~A.~Samuel and D.~Sahdev,
\PRL 43 746 1979 ;\nextline
C.~L.~Bilchak, R.~W.Brown and J.~D.~Stroughair, \PR D29 375 1984 ;\nextline
J.~Cortes, K.~Hagiwara and F.~Herzog, \NP B278 26 1986 ;\nextline
U.~Baur and D.~Zeppenfeld, \NP B308 127 1988 .}
\REF\HHPZ{K.~Hagiwara \etal, \NP B282 253 1987 ;\nextline
D.~Zeppenfeld, \PL 183B 380 1987 .}
\REF\WZ{S.~Willenbrock and D.~Zeppenfeld, \PR D37 1775 1988 ;\nextline
K.~Hagiwara, J.~Woodside and D.~Zeppenfeld, \PR D41 2113 1990 .}
\REF\WEP{E.~Gabrielli, Mod. Phys. Lett. {\bf A1}, 465 (1986);\nextline
M.~B\"ohm and A.~Rosado, \ZP C42 479 1989 .}
\REF\Epw{U.~Baur and D.~Zeppenfeld, \NP B325 253 1989 .}
\REF\GG{E.~Yehudai, \PR D41 33 1990 ; \PR D44 3434 1991 ;\nextline
S.~Y.~Choi and F.~Schrempp, \PL 272B 149 1991 .}
\REF\Arg{E.~N.~Argyres \etal\ , DTP/91/88 preprint, (December 1991).}
\REF\Pet{J.~Alitti \etal\ (UA2 Collaboration), CERN-PPE/91-216 preprint
(December 1991).}
\REF\LHC{The LHC Study Group, Design Study of the Large Hadron Collider,
CERN 91-03, 1991.}
\REF\God{S.~Godfrey, OCIP/C 91-2 preprint (May 1991).}
\REF\HS{T.~Helbig and H.~Spiesberger, DESY 91-096 preprint (September
1991).}
\REF\Jog{J.~M.~Cornwall, D.~N.~Levin and G.~Tiktopoulos,
\PRL 30 1268 1973 ; \PR D10 1145 1974 ;\nextline
C.~H.~Llewellyn Smith, \PL 46B 233 1973 ;\nextline
S. D. Joglekar, Ann. of Phys. {\bf 83} 427 (1974).}
\REF\BZ{U.~Baur and D.~Zeppenfeld, \PL 201B 383 1988 .}
\REF\HZ{K. Hagiwara and D. Zeppenfeld, \NP B274 1 1986 .}
\REF\GoG{K.~Gaemers and G.~Gounaris, \ZP C1 259 1979 .}
\REF\Hagi{W.~J.~Marciano and A.~Queijeiro, \PR D33 3449 1986 ;\nextline
F.~Boudjema, K.~Hagiwara, C.~Hamzaoui and K.~Numata,
\PR D43 2223 1991 .}
\REF\HMRS{P.~N.~Harriman, A.~D.~Martin, R.~G.~Roberts and
W.~J.~Stirling, \PR D42 798 1990 .}
\REF\Higgs{R.~N.~Cahn, S.~D.~Ellis, R.~Kleiss and W.~J.~Stirling,
\PR D35 1626 1987 ;\nextline
R.~Kleiss and W.~J.~Stirling, \PL 200B 193 1988 ;\nextline
U.~Baur and E.~W.~N.~Glover, \NP B347 12 1990 ;\nextline
V.~Barger \etal\ , \PR D44 2701 1991 .}
\REF\BB{U.~Baur and E.~L.~Berger, \PR D41 1476 1990 .}
%
%
\FIG\one{Distribution of a) the jet photon separation,
$\Delta R_{j\gamma}$, and b) the jet photon invariant mass,
$m_{j\gamma}$, for
$e^-p\to\gamma j\psl_T$ (solid line) and $e^+p\to\gamma j\psl_T$ (dashed
line) at HERA in the SM. The cuts imposed are specified in the text. }
\FIG\two{Transverse momentum distribution of the photon in a)
$e^-p\to\gamma j\psl_T$ and b) $e^+p\to\gamma j\psl_T$ at HERA for the SM
(solid line) and various anomalous values of $\kappa$ and $\lambda$.
Cuts are specified in the text.}
\FIG\three{Transverse momentum distribution of the photon in a)
$e^-p\to\gamma j\psl_T$ and b) $e^+p\to\gamma j\psl_T$ at LEP/LHC for the SM
(solid line) and various anomalous values of $\kappa$ and $\lambda$.
Cuts are specified in the text.}
\FIG\four{Jet pseudorapidity distribution for $e^-p\to\gamma j\psl_T$ at
LEP/LHC in the region $p_{T\gamma}>200$~GeV. The solid line displays the
SM result, whereas the dashed, dash-dotted and dotted curves show the
predictions for $\Delta\kappa=-1$, $\Delta\kappa=+1$ and $\lambda=0.1$. }
%
\pagenumber=1

One of the prime targets for experiments at present and future colliders
is the measurement of the $WW\gamma$ and $WWZ$ couplings. In the
Standard Model (SM) of electroweak interactions, these couplings are
unambiguously fixed by the nonabelian nature of the $SU(2)\times U(1)$
gauge symmetry. In contrast to low energy
and high precision experiments at the $Z$ peak, collider experiments
offer the possibility of a direct, and essentially model independent,
measurement of the three vector boson vertices. In the past, a number of
different collider processes which are
sensitive to anomalous $WW\gamma$ and $WWZ$ couplings have been
studied (see \eg\ Refs.~\Sam~--~\Arg). Analyzing the reaction
$p\bar p\to e^\pm\nu\gamma X$, the UA2 Collaboration recently reported
the first direct measurement of the $WW\gamma$ vertex\rlap.
\refmark{\Pet} More precise information on anomalous $WW\gamma$ couplings
can soon be expected from Tevatron experiments, as well as from HERA.

At $ep$ colliders such as HERA (30~GeV electrons/positrons on 820~GeV
protons; $\sqrt{s}=314$~GeV, ${\cal L}=2\cdot 10^{31}{\rm cm}^{-2}{\rm
s}^{-1}$) or LEP/LHC (60~GeV electrons/positrons on 7.7~TeV protons;
$\sqrt{s}=1.36$~TeV, ${\cal L}=2.8\cdot
10^{32}{\rm cm}^{-2}{\rm s}^{-1}$\refmark{\LHC}) single $W$ boson
production via $ep\to eW^\pm X$ and radiative charged current
scattering, $ep\to\nu\gamma X$,
offer chances to test the $WW\gamma$ vertex. Single $W$ boson production
in $ep$ collisions has been studied extensively in the
past\rlap.\refmark{\WEP,\Epw}
The process $e^-p\to\nu\gamma X$ has only recently been investigated, with
conflicting results reported in Refs.~\God\ and~\HS.
In this paper we present an independent calculation of the
process $e^-p\to\nu\gamma X$, using the most general $WW\gamma$
vertex compatible with Lorentz and electromagnetic gauge invariance. We
also extend the existing studies, investigating the reaction
$e^+p\to\nu\gamma X$, and take into account the form factor behaviour
of the nonstandard $WW\gamma$ couplings required by $S$-matrix
unitarity\rlap.\refmark{\Jog,\BZ}

The process $ep\to\nu\gamma X$ offers potential
advantages over $eW$ production in measuring the $WW\gamma$ vertex. At
HERA, and to a smaller degree at
LEP/LHC, $W$ production proceeds close to the kinematical threshold;
furthermore, only the leptonic decays of the $W$ boson can be identified.
Both factors reduce the number of events which can be utilized in
analyzing the structure of the $WW\gamma$ vertex in $ep\to eWX$. On the
other hand,
$ep\to\nu\gamma X$ is a pure charged current process which is suppressed
by the large mass of the exchanged $W$. It is thus not clear {\sl a priori}
whether $ep\to\nu\gamma X$ or $ep\to eWX$ will be more sensitive to
anomalous $WW\gamma$ couplings.

There are four Feynman diagrams contributing to the reaction $e^\pm
p\to\nu\gamma X$ at the parton level. The photon can be either radiated
from the incoming
lepton or quark line, from the outgoing quark line, or from the
exchanged $W$ boson. Using the spinor
technique described in Ref.~\HZ, we have calculated the matrix elements
for $e^\pm q\to\nu\gamma q'$ for the most general $WW\gamma$ vertex
compatible with Lorentz and electromagnetic gauge invariance. Since the
exchanged $W$ couples to essentially massless quarks which effectively
insures that $\partial_\mu W^\mu=0$, the $WW\gamma$ vertex depends on
four free parameters only, and can conveniently be described by the
effective Lagrangian\rlap,\refmark{\HHPZ,\GoG}
$$\eqalign{{\cal L}_{WW\gamma} &= -ie \left\{ \left( W^{\dag}
_{\mu\nu} W^\mu A^\nu - W^{\dag}_\mu A_\nu W^{\mu\nu} \right)
+ \kappa W^{\dag}_\mu W_\nu F^{\mu\nu} +
{\lambda \over M^2_W} W_{\lambda\mu}^{\dag} W^\mu_{ \ \nu} A^{\nu\lambda}
\right. \crr \eqinsert{(1)}
& \qquad \qquad \qquad \qquad \qquad
\left. +\tilde\kappa W^{\dag}_\mu W_\nu \tilde F^{\mu\nu} +
{\tilde\lambda \over M^2_W} W^{\dag}_{\lambda\mu} W^\mu_{ \ \nu}
\tilde F^{\nu\lambda} \right\} . \cr} $$
Here $A^\mu$ and $W^\mu$ are the photon and $W^-$ fields, respectively,
$W_{\mu\nu} = \partial_\mu W_\nu - \partial_\nu W_\mu, \ F_{\mu\nu} = \partial
_\mu A_\nu - \partial_\nu A_\mu$, and $\tilde F_{\mu\nu} = {1\over 2}
\epsilon_{\mu\nu\rho\sigma} F^{\rho\sigma}$. $e$ is the charge of the
proton, and $M_W$ the mass of the $W$ boson.

The first term in Eq.~(1) arises from the minimal coupling of the photon to
the $W^\pm$ fields and is completely fixed by the charge of the $W$ boson
for onshell photons.  The $\kappa$ and $\lambda$ terms are related to the
magnetic dipole moment $\mu_W$ and the electric quadrupole moment $Q_W$
of the $W^+$
$$\mu_W = {e\over 2M_W} \left( 1 + \kappa + \lambda \right)\>,
\hskip 1.cm
Q_W = -{e\over M^2_W}\, (\kappa -\lambda) \>. \eqno(2) $$
Within the SM, at tree level,
$$\kappa =1 \hskip 1.cm {\rm and} \hskip 1.cm \lambda =0 \>. \eqno(3) $$
The $CP$ violating couplings $\tilde\kappa$ and $\tilde\lambda$, which
both vanish at tree level in the SM, are constrained by the electric
dipole moment of the neutron to be smaller than ${\cal O}(10^{-3})
$\refmark{\Hagi} in magnitude. Subsequently we shall therefore concentrate on
the anomalous couplings $\kappa$ and $\lambda$.

Tree level unitarity restricts the $WW\gamma$ couplings to their (SM)
gauge theory values at asymptotically high energies\rlap.\refmark{\Jog,\BZ}
This implies that any deviation $a=\Delta\kappa=\kappa-1,~\lambda$ from
the SM expectation has to be described by a form factor $a(q^2_W,
\bar q^2_W,q^2_\gamma=0)$, which vanishes when either $|q^2_W|$, or
$|\bar q^2_W|$, the absolute square of the four-momentum of the
exchanged $W$ bosons, becomes large. Consequently, we shall include
form factors
$$ a(q^2_W,\bar q^2_W,0)=a_0\left[\left(1-{q^2_W\over\Lambda^2}\right)
\left(1-{\bar q^2_W\over\Lambda^2}\right)\right]^{-n} \eqno(4) $$
with $n=1$ in all our calculations. The scale $\Lambda$ in Eq.~(4) represents
the
scale at which new physics becomes important in the weak boson sector,
\eg\ due to a composite structure of the $W$ boson. We shall use
$\Lambda=1$~TeV in our numerical simulations.

We have compared our squared $e^-p\to\nu\gamma X$
matrix elements in the limit $\Lambda\to\infty$ with those of Ref.~\HS\
for arbitrary values of $\kappa$ and $\lambda$.
The numerical agreement is excellent. On the other hand, using the same set of
parameters and cuts as in Ref.~\God, we are unable to reproduce the
cross section values
and the photon transverse momentum distribution of Ref.~\God.

Subsequently we shall discuss $e^-p\to\nu\gamma X$ and
$e^+p\to\nu\gamma X$ in parallel.
The SM parameters which will be used in all figures and tables are
$M_W=80$~GeV and $\sin^2\theta_W=0.23$. The cross section for $e^\pm
p\to\nu\gamma X$ is proportional to $\alpha^3$, where $\alpha$ is the
electromagnetic coupling constant. Since a real photon is emitted, one
factor is evaluated at scale $m_e^2$ where $m_e$ is the electron mass
($\alpha(m_e^2)=1/137$), and the two remaining factors are taken as
$\alpha(M_W^2)=1/128$. For the proton
structure functions we use the HMRSB set\refmark{\HMRS} with the scale
$Q^2$ given by the four-momentum transfer to the scattered quark.
Uncertainties in the energy measurements of the photon and jet in the
detector are taken into account by Gaussian smearing of the four momenta
with standard deviation $\sigma=(0.15~{\rm GeV}^{1/2})\sqrt{E}$ and
$\sigma=(0.35~{\rm GeV}^{1/2})\sqrt{E}$, respectively.

The signal we are investigating consists of a photon, missing transverse
momentum, $\psl_T$, which originates from the neutrino, and a jet
produced by the quark struck inside the proton. In order to regulate
the infrared and collinear singularities present in
$e^\pm p\to\nu\gamma X$, it is necessary to impose a nonzero cut on the
photon transverse momentum $p_{T\gamma}$ and the jet photon separation
$\Delta R_{j\gamma}=\left[(\Delta\phi_{j\gamma})^2+(\Delta\eta_{j\gamma}
)^2\right]^{1/2}$ in the pseudorapidity-azimuthal angle plane.

At HERA energies, the $\gamma j\psl_T$ final state is predominantly produced
from a valence $u$-quark ($d$-quark) in the proton for $e^-p$ ($e^+p$)
collisions. Photons in the collinear region therefore are radiated mostly from
(final state) $d$ ($u$) quarks in $e^-p\to\gamma j\psl_T$
($e^+p\to\gamma j\psl_T$). Due to the larger electric charge of the
$u$-quarks, the collinear singularity thus is expected to be considerably more
pronounced in the $e^+p$ case. On the other hand, the difference in the $u$
and $d$ valence quark distributions in the proton tends to suppress the
$e^+p\to\gamma j\psl_T$ cross section.

Both effects are clearly reflected in the distributions of the jet photon
separation $\Delta R_{j\gamma}$ and the invariant mass
$m_{j\gamma}$, shown in Fig.~1. To roughly simulate
the finite acceptance of detectors, we require $p_{T\gamma}>5$~GeV, a missing
transverse momentum and a jet $p_T$ of $\psl_T,~p_{Tj}>10$~GeV,
and impose a photon and jet rapidity cut of $|\eta_\gamma|,|\eta_j|<3.5$.
At large $\Delta R_{j\gamma}$ ($m_{j\gamma}$), the rate for
$e^+p\to\gamma j\psl_T$ is suppressed. In the collinear region,
however, the $e^+p\to\gamma j\psl_T$ cross section rises much faster
than the $e^-p\to\gamma j\psl_T$ rate, and the two cross sections are
very similar. The peak at $\Delta R_{j\gamma}\approx\pi$ and
$m_{j\gamma}\approx 30$~GeV arises from photons which are radiated from
the incoming $e^\pm$ line. The $j\gamma$ invariant mass distribution
shows a rather long tail, extending out to about one half of the
available center of mass energy.

Due to the infrared singularity associated with photon emission from the
incoming quark and lepton line, the transverse momentum distribution of
the photon strongly peaks at small $p_{T\gamma}$ values. Since the
$WW\gamma$ vertex does not enter those diagrams, the signal at low
$p_{T\gamma}$ is insensitive to anomalous $WW\gamma$ couplings, and no
sensitivity is lost by requiring a hard photon with $p_{T\gamma}>10$~GeV
(20~GeV) at HERA (LEP/LHC). The
transverse momentum distribution of the photon in $e^\pm p\to\gamma j\psl_T$
at HERA is shown in Fig.~2 for the SM (solid line) and anomalous values of
$\kappa$ and $\lambda$. Fig.~3 displays $d\sigma/dp_{T\gamma}$ at LEP/LHC.  We
have imposed a jet and missing $p_T$ cut of $p_{Tj},~\psl_T>10$~GeV (20~GeV)
at HERA (LEP/LHC), and a jet pseudorapidity cut of $|\eta_j|<3.5$ (4.5).
Furthermore we have required $|\eta_\gamma|<3.5$ and the jet and photon to be
well separated, $\Delta R_{j\gamma}>0.5$.

{}From Figs.~2 and~3 it is obvious that $e^\pm p\to\gamma j\psl_T$ is
sensitive to anomalous $WW\gamma$ couplings at large values of
$p_{T\gamma}$. However, the Feynman diagram involving the $WW\gamma$
vertex contains two $W$ propagators and thus is suppressed with respect
to the bremsstrahlung diagrams. As a result, at HERA, rather large anomalous
couplings are necessary to produce significant deviations from the SM
$p_{T\gamma}$ distribution. Figs.~2b and~3b show that the photon transverse
momentum distribution for $e^+p\to\gamma j\psl_T$ falls considerably
faster than for $e^-p\to\gamma j\psl_T$ in the SM. Furthermore, for a
given (non-standard) value of $\kappa$ or $\lambda$, deviations from the
SM prediction are larger in $e^+p\to\gamma j\psl_T$. Whereas the
form factor behaviour negligibly influences predictions for HERA, cross
sections are reduced by about a factor~2 at LEP/LHC for anomalous
$WW\gamma$ couplings and large photon transverse momenta.

At HERA as well as LEP/LHC energies the sensitivity to anomalous
couplings in $e^\pm p\to\gamma j\psl_T$ effectively stems from regions
in phase space where the anomalous contributions to the cross section
are smaller than the SM expectation. One therefore expects that
interference effects between the SM amplitude and the anomalous
contributions to the amplitude play a non-negligible role. This effect
is most pronounced for anomalous values of $\kappa$. At intermediate
photon transverse momenta, destructive interference causes
$d\sigma/dp_{T\gamma}$ to dip below the SM prediction for positive
values of $\Delta\kappa$. At large values of $p_{T\gamma}$, the
interference term changes sign and significantly reduces the sensitivity
to negative values of $\Delta\kappa$. In contrast to the naive expectation,
deviations from the SM prediction do not grow
with increasing $p_{T\gamma}$ at HERA and LEP/LHC energies for
$\Delta\kappa<0$ (provided that $|\Delta\kappa|$ is not too large).

This effect is enhanced, in particular in the $e^-p$ case, by the
finite rapidity cut imposed on the jet in
Figs.~2 and~3. For large values of the
photon transverse momentum and negative $\Delta\kappa$, the jet has a rapidity
distribution which extends to considerably larger (negative) rapidities
than for the SM, or other anomalous $WW\gamma$ couplings. This is
demonstrated in Fig.~4, where we show the jet pseudorapidity
distribution for $e^-p\to\gamma j\psl_T$ at LEP/LHC in the region
$p_{T\gamma}>200$~GeV in the SM case (solid line), for $\Delta\kappa=\pm
1$, and for $\lambda=0.1$. With the exception of $\eta_j$ and
$p_{T\gamma}$, all other cuts are as in Fig.~3. The incoming proton is
assumed to move
in the {\sl negative} $z$ direction. For $\Delta\kappa=-1$, the $\eta_j$
distribution peaks at a somewhat larger value than in the other cases,
and a substantial portion of the cross section originates from the
region $\eta_j<-3.5$.

The $\eta_j$ distribution directly reflects the properties of the
anomalous contributions to the helicity amplitudes. For non-standard
values of $\kappa$, the photon mostly couples to longitudinally
polarized $W$'s at high energies\rlap.\refmark{\BZ}
The situation in $e^\pm p\to\gamma j\psl_T$ thus is similar to that of
heavy Higgs boson
production via vector boson fusion at hadron colliders. Heavy Higgs
bosons mostly couple to longitudinal vector bosons, which causes the
jet rapidity distribution in $qq\to qqH$ to peak at large $\eta_j$\rlap.
\refmark{\Higgs} Indeed, in the region $\eta_j\lsim -4.5$ and at large
$p_{T\gamma}$, the jet rapidity
distributions for $\Delta\kappa=+1$ and $\Delta\kappa=-1$ in
$e^-p\to\gamma j\psl_T$ are very
similar, indicating that the anomalous contribution dominates in this
region. In the more central rapidity region, the largest contribution to the
cross section originates from the interference term, which tends to
cancel against the SM contribution for negative $\Delta\kappa$.

As mentioned above, rather large anomalous couplings are
necessary in order to produce measurable effects at HERA. This
qualitative statement can be made more quantitative by deriving those values
of $\kappa$ and $\lambda$ which would give rise to a deviation from the
SM at the 90\% and 69\% confidence level
(CL) in the $p_{T\gamma}$ spectrum for an integrated luminosity of
$\int\!{\cal L}dt=10^3~{\rm pb}^{-1}$, calculated within the cuts
specified above. At HERA, an integrated luminosity of 1000~pb$^{-1}$
corresponds to at least five years of running.
The confidence level is calculated
by splitting the $p_{T\gamma}$ distribution into 6 (10) bins at HERA
(LEP/LHC) for $e^-p\to\gamma j\psl_T$, and into 4 (8) bins for
$e^+p\to\gamma j\psl_T$. The last $p_{T\gamma}$ bin contains all events above
a certain threshold, in order to achieve a sizeable counting rate
(more than 5~events) in each bin.  This procedure guarantees that in our
calculation a high confidence level cannot arise from a single event at
high $p_T$ where the SM predicts, say, only 0.01 events. In order to
derive realistic limits we allow for a normalization uncertainty of the
SM cross section of $\Delta{\cal N}=30\%$. No attempt has been made to
take into account the possible change in the shape of the $p_{T\gamma}$
distribution due to higher order QCD corrections.

To illustrate the sensitivities which can be achieved, we list
the minimal anomalous couplings, which would give rise to a
90\% or 69\% CL effect, in Table~1 for the case where only one coupling
at a time
is assumed to be different from the SM value. Although the effects of
anomalous $WW\gamma$ couplings on the photon transverse momentum distribution
are more pronounced in $e^+p\to\gamma j\psl_T$, the smaller event rate,
combined with the faster falling
$p_{T\gamma}$ distribution, in general results in limits which are
significantly weaker than those for $e^-p\to\gamma j\psl_T$ at HERA.
Only for $\Delta\kappa<0$ are the sensitivity limits for
$e^+p\to\gamma j\psl_T$ and $e^-p\to\gamma j\psl_T$ comparable. Even at
LEP/LHC the bounds for $e^-p$ collisions are slightly better. The limits
shown in Table~1 are somewhat weaker than those presented
in Ref.~\HS, mostly as a result of the possible normalization
uncertainty in the SM prediction which we included in our analysis.
Because interference terms, \ie\ terms linear in $\Delta\kappa$
and $\lambda$, dominate over those quadratic in the anomalous couplings,
the bounds scale essentially with the square root of the integrated luminosity.

Finally, we compare the limits achievable in $e^\pm p\to\gamma j\psl_T$
with bounds from $S$-matrix unitarity and the sensitivity to non-gauge
theory terms in the $WW\gamma$ vertex
accessible in other present and future collider experiments. Bounds from
$S$-matrix unitarity depend explicitly on the functional form and the
scale $\Lambda$ of the form factor. Varying only one coupling at a time,
the following upper limits are obtained from unitarity for the form
factor of Eq.~(4) and $\Lambda\gg M_W$:
$$ n=1:\cases{ |\Delta\kappa|\!\!\!&$\lsim 1.9~{\rm TeV}^2/\Lambda^2$\cr
\phantom{\Delta}|\lambda|\!\!\!&$\lsim 1.0~{\rm TeV}^2/\Lambda^2$}
\hskip 1.cm
n=2:\cases{ |\Delta\kappa|\!\!\!&$\lsim 7.6~{\rm TeV}^2/\Lambda^2$\cr
\phantom{\Delta}|\lambda|\!\!\!&$\lsim 4.0~{\rm TeV}^2/\Lambda^2$}
\eqno (5) $$
Comparing the bounds listed in Table~1 with Eq.~(5), one observes that
measuring the anomalous $WW\gamma$ couplings in either $e^- p\to\gamma
j\psl_T$ or $e^+p\to\gamma j\psl_T$ at HERA will not significantly
improve the limits from $S$-matrix unitarity.

As we mentioned at the beginning, experiments at HERA, studying $eW$
production, will also be able to probe the $WW\gamma$ vertex. For the
same integrated luminosity used to derive the limits of Table~1, the
bounds achievable in $ep\to eWX$\refmark{\Epw} are about a factor~2
to~3 better than those from $e^\pm p\to\gamma j\psl_T$. Even at LEP/LHC $eW$
production will be somewhat more sensitive. The UA2 Collaboration
has recently measured $\kappa$ and $\lambda$ in the process $p\bar p\to
e^\pm\nu\gamma X$ at the CERN $p\bar p$ collider, obtaining\refmark{\Pet}
$$ \kappa=1\matrix{+2.6\crc -2.2}~~({\rm for}~\lambda=0)\hskip 1.
cm\lambda=0\matrix{+1.7\crc -1.8}~~({\rm for}~\kappa=1). \eqno (6) $$
The errors in Eq.~(6)
are already within a factor of two of, or even better than those which can
be expected from $e^\pm p\to\gamma j\psl_T$ with 1000~pb$^{-1}$ at HERA.
Moreover, they are expected to be reduced considerably in the near
future with new Tevatron data. With an integrated luminosity of $100~{\rm
pb}^{-1}$, $|\Delta\kappa|$ can be constrained to be less than
0.7~--~1.0 (1.1~--~1.5)
at 69\% (90\%) CL in $p\bar p\to W^\pm\gamma$, whereas $|\lambda|$ can be
measured to $|\lambda|<0.25$~--~0.30 (0.40~--~0.50)\rlap.\refmark{\BB}
An even more precise determination of the
anomalous $WW\gamma$ couplings will be possible in $e^+e^-\to W^+W^-$ at
LEP~II where an accuracy of $|\Delta\kappa|$, $|\lambda|\approx 0.1-0.2$
is expected\rlap.\refmark{\HHPZ}

In summary, we have presented an independent calculation of the process
$e^- p\to\nu\gamma X$, using the most general $WW\gamma$ vertex
allowed by Lorentz and electromagnetic gauge invariance. Our
matrix elements fully agree with those of Ref.~\HS,
while we are unable to reproduce the numerical results presented in
Ref.~\God. We also explored the possibilities of probing the $WW\gamma$
vertex in the reaction $e^+p\to\nu\gamma X$.
Although the effects of anomalous $WW\gamma$ couplings on the
photon transverse momentum distribution are more pronounced in
$e^+p\to\gamma j\psl_T$, the smaller event rate, in particular at large
photon transverse momenta, severely limits the bounds on $\Delta\kappa$
and $\lambda$ which can be achieved. We found that they are, at best,
comparable to those obtained in $e^-p\to\nu\gamma X$. In the energy
domain of HERA and LEP/LHC, destructive interference
effects between the SM and anomalous contributions to the amplitude, combined
with effects induced by the finite jet pseudorapidity coverage of detectors,
result in sensitivity bounds for $e^\pm p\to\nu\gamma X$, which are
significantly weaker than those which can be expected from $ep\to eWX$,
$p\bar p\to e^\pm\nu\gamma X$, and $e^+e^-\to W^+W^-$ within the next few
years. At HERA, the limits which can be achieved for $\kappa$ and
$\lambda$ in $e^\pm p\to\nu\gamma X$ are similar to the bounds resulting
from $S$-matrix unitarity.
\vskip .35in
\ack
We would like to thank W.~Buchm\"uller, W.~Smith, D.~Zeppenfeld and
P.~Zerwas for useful discussions. We are also
grateful to H.~Spiesberger for useful correspondence and his
assistance in comparing our results with those of Ref.~\HS. This work
was supported in part by the U.~S. Department of Energy under Contract
No.~DE-AC02-76ER00881 and Contract No.~DE-FG05-87ER40319, in part by the
Texas National Research Laboratory Commission under Grant No.~RGFY9173,
and in part by the University of Wisconsin Research Committee with funds
granted by the Wisconsin Alumni Research Foundation.

\endpage
\refout
\endpage
\centerline{TABLE~1}
\vskip 3.mm
\noindent
Sensitivities achievable at the 90\% and 69\% CL for the anomalous
$WW\gamma$ couplings $\Delta\kappa=\kappa-1$ and $\lambda$ in
$e^-p\to\gamma j\psl_T$ and $e^+p\to\gamma j\psl_T$ at HERA and LEP/LHC for an
integrated luminosity of 1000~pb$^{-1}$. Only one coupling at a time is
assumed to be different from the SM value.
\vskip 0.5in
\tablewidth=5.5in
\def\tstrut{\vrule height 4.3ex depth 2.5ex width 0pt}
\begintable
coupling & CL & HERA & LEP/LHC \nr
\multispan{4}\tstrut\hfil $e^-p\to\gamma j\psl_T$ \hfil\crthick
$\Delta\kappa$  | 90\% | $\matrix{+2.2 \crc -2.4}$ | $\matrix{+0.44\crc
-0.54}$ \nr
 | 69\% | $\matrix{+1.3 \crc -1.4}$ | $\matrix{+0.25 \crc -0.30}$ \cr
$\lambda$  | 90\% | $\matrix{+2.8 \crc -2.1}$ | $\matrix{+0.17\crc -0.12}$ \nr
 | 69\% | $\matrix{+2.1 \crc -1.3}$ | $\matrix{+0.12 \crc -0.08}$
 \crthick
\multispan{4}\tstrut\hfil $e^+p\to\gamma j\psl_T$ \hfil\crthick
$\Delta\kappa$  | 90\% | $\matrix{+4.0 \crc -2.2}$ | $\matrix{+0.53\crc
-0.53}$ \nr
 | 69\% | $\matrix{+3.0 \crc -1.1}$ | $\matrix{+0.31 \crc -0.30}$ \cr
$\lambda$  | 90\% | $\matrix{+4.8 \crc -3.8}$ | $\matrix{+0.21\crc -0.16}$ \nr
 | 69\% | $\matrix{+3.1 \crc -2.3}$ | $\matrix{+0.16 \crc -0.10}$
\endtable
\endpage
\figout
\bye